\label{key}
\documentclass[twocolumn,aps,superscriptaddress,nofootinbib,floatfix,preprintnumbers,amsmath,amssymb]{revtex4}

\usepackage{epsfig,bm}
\usepackage{graphics}
\usepackage{color}
\usepackage{slashed}

\begin{document}


\title{Overall momentum balance and redistribution of the lost energy in asymmetric dijet events in 2.76~ATeV Pb-Pb collisions with a multi-phase transport  model}


\author{Zhan Gao}
\affiliation{Key Laboratory of Quark and Lepton Physics (MOE) and Institute of Particle Physics, Central China Normal University, Wuhan 430079, China}

\author{Ao Luo}
\affiliation{Key Laboratory of Quark and Lepton Physics (MOE) and Institute of Particle Physics, Central China Normal University, Wuhan 430079, China}

\author{Guo-Liang Ma}
\affiliation{Shanghai Institute of Applied Physics, Chinese Academy of Sciences, Shanghai 201800, China}

\author{Guang-You Qin}
\affiliation{Key Laboratory of Quark and Lepton Physics (MOE) and Institute of Particle Physics, Central China Normal University, Wuhan 430079, China}

\author{Han-Zhong Zhang}
\affiliation{Key Laboratory of Quark and Lepton Physics (MOE) and Institute of Particle Physics, Central China Normal University, Wuhan 430079, China}


\date{\today}

\begin{abstract}

The overall transverse momentum balance and the redistribution of the lost energy from hard jets for asymmetric dijet events in PbPb collisions at 2.76~ATeV at the LHC is studied within A Multi-Phase Transport (AMPT) model.
A detailed analysis is performed for the projected transverse momentum $\langle \slashed{p}_{T}^{||} \rangle$ contributed from the final charged hadrons carrying different transverse momenta and emitted from different angular directions.
We find that the transverse momentum projection $\langle \slashed{p}_{T}^{||} \rangle $ in the leading jet direction is mainly contributed by hard hadrons ($p_T > 8.0$~GeV/$c$) in both peripheral and central PbPb collisions, while the opposite direction in central collisions is dominated by soft hadrons ($p_T = 0.5$-$2.0$~GeV/$c$).
The study of in-cone and out-of-cone contributions to $\langle \slashed{p}_{T}^{||} \rangle$ shows that these soft hadrons are mostly emitted at large angles away from the dijet axis.
Our AMPT calculation is in qualitative agreement with the CMS measurements and the primary mechanism for the energy transported to large angles in the AMPT model is the elastic scattering at the partonic stage.
Future studies including also inelastic processes should be helpful in understanding the overestimation of the magnitudes of in-cone and out-of-cone imbalances from our AMPT calculations, and shed light on different roles played by radiative and collisional processes in the redistribution of the lost energy from hard jets.

\end{abstract}


\maketitle

\section{Introduction}
\label{sec:intro}

Jet quenching provides very important evidence for the formation of the hot and dense quark-gluon plasma (QGP) in high-energy heavy-ion collisions  \cite{Wang:1991xy, Qin:2015srf}.
It originates from the energy loss experienced by the hard partonic jets initially produced from early scatterings as they traverse and interact with the highly excited nuclear matter created in these energetic collisions.
The picture of parton energy loss and jet quenching has been confirmed by a wealth of experimental results observed at the Relativistic Heavy-Ion Collider (RHIC) and the Large Hadron Collider (LHC), such as the suppression of large transverse momentum hadron production \cite{Adams:2003im,Adams:2003kv,Adler:2003qi,CMS:2012aa, Abelev:2012hxa}, and the strong modification of dihadron and photon-hadron transverse momenta and azimuthal angle correlations \cite{Adler:2002tq,Aamodt:2011vg,Adare:2009vd,Abelev:2009gu}, in central nucleus-nucleus collisions as compared to elementary nucleon-nucleon collisions.
Various theoretical and phenomenological models have been developed to explain these jet modification phenomena \cite{Vitev:2002pf,Wang:2003mm,Eskola:2004cr,Renk:2006pk,Zhang:2007ja,Qin:2007rn,Bass:2008rv,Zhang:2009rn,Qin:2009bk,Chen:2011vt,Majumder:2011uk,Zapp:2012ak,Burke:2013yra,Andres:2016iys,Cao:2013ita}, and the comparisons of theories to experimental data have shown that jet energy loss is due to the combined effects of elastic and inelastic interactions between the propagating hard partons and the constituents of the hot and dense QGP matter.

In recent years, much attention has been devoted to fully reconstructed jet observables in relativistic heavy-ion collisions.
As full jets include the contributions from both leading and subleading fragments of the parton showers, they are expected to provide more detailed information than hadronic observables on the interaction between jet and medium.
Various full jet observables have been studied in heavy-ion experiments at RHIC and the LHC, e.g., single inclusive full jet spectra \cite{Salur:2008hs, Putschke:2008wn,Bruna:2009em, Ploskon:2009zd, Aad:2014bxa}, the transverse momentum asymmetry distributions and angular correlations for dijet and photon-jet events \cite{Chatrchyan:2011sx,Aad:2010bu, Jacobs:2015srw, Adam:2015doa}, and the internal structures of the full jets \cite{Chatrchyan:2012gw,Chatrchyan:2013kwa,Aad:2014wha}.
In order to understand the observed nuclear modifications of full jet production and structure, it is required to develop theoretical models and calculations that include the effect of the medium on both leading and subleading partons of the full jets \cite{Qin:2010mn,CasalderreySolana:2010eh,Vitev:2009rd,Young:2011qx,He:2011pd,Dai:2012am,Qin:2012gp,Wang:2013cia,Ma:2013pha,CasalderreySolana:2012ef,Blaizot:2013hx,Fister:2014zxa,Chien:2015hda,Milhano:2015mng,Chang:2016gjp,Casalderrey-Solana:2016jvj,Mueller:2016gko,Mueller:2016xoc,Chen:2016vem}.
The comparisons between theories and experiments have demonstrated that full jets may experience a significant amount of energy loss as well when they propagate through the hot and dense nuclear matter, and the distribution of the energy and momentum inside full jets may also be strongly modified by the interaction with the medium constituents.

In addition to parton energy loss (jet quenching), the other important aspect of jet-medium interaction is the medium response: when hard partonic jets propagate through the hot and dense QGP matter, they not only lose energy due to jet-medium interaction, but also induce medium excitation \cite{CasalderreySolana:2004qm, Chaudhuri:2005vc, Ruppert:2005uz, Qin:2009uh, Neufeld:2009ep,Tachibana:2017syd}.
In particular, the energy and momentum lost from hard jets are deposited into the medium, which may modify the subsequent medium evolution and manifest in the final-state hadron distributions and correlations.
To investigate where the lost energy from the jets appears in the final states, CMS Collaboration has measured the so-called projected transverse momentum $\langle \slashed{p}_{T}^{||} \rangle$ for asymmetric dijet events, and the experimental analysis tends to indicate that a large portion of the the lost energy from jets is carried by the soft hadrons emitted at large angles away from the dijet propagation direction \cite{CMS:2014uca}.

In this paper, we simulate the evolution and the redistribution of the lost energy from the hard jets using the framework of A Multi-Phase Transport Model (AMPT) \cite{Lin:2004en}.
In particular, we follow the CMS collaboration and study the so-called projected transverse momentum $\langle \slashed{p}_{T}^{||} \rangle$ for asymmetric dijet events.
It is worth noting that in Ref.~\cite{Tachibana:2014lja} this observable has been studied with a (3+1)-dimensional hydrodynamic model, using a simplified energy deposition profile.
Here we simulate both jet propagation and medium evolution simultaneously with the AMPT model, and perform a detailed analysis for the various contributions to $\langle \slashed{p}_{T}^{||} \rangle$ from the final state hadrons carrying different transverse momenta and emitted from different angular directions with respect to the dijet propagation direction.
Our simulation results are qualitatively consistent with the CMS observation which shows that a large portion of the deposited energy and momentum by the hard partonic jets is transported by elastic collisions (in the AMPT model) and finally carried by the soft hadrons emitted at large angles away from the dijet propagation direction.

The paper is organized as follows.
In Sec. II, we provide a brief introduction to the AMPT model and the corresponding settings used in our studies.
The numerical results for the transverse momentum projection $\langle \slashed{p}_{T}^{||} \rangle$ for asymmetric dijet events are presented and discussed in detail in Sec. III.
Sec. IV contains the summary.

\section{The AMPT Model}
\label{sec:model}

In this work, we use the AMPT model with string melting mechanism~\cite{Lin:2004en}, which has provided good descriptions of various soft bulk observables at the LHC energies~\cite{Xu:2011fi}. In addition, the AMPT model with a triggered dijet can also describe many aspects of full reconstructed jet observables, such as the transverse momentum $p_{T}$ asymmetry of dijet or photon-jet events~\cite{Ma:2013pha, Ma:2013bia}, jet fragmentation function~\cite{Ma:2013gga}, jet shape function~\cite{Ma:2013uqa} and jet anisotropy parameter~\cite{Nie:2014pla}.

There are four main stages in the AMPT model to simulate high-energy heavy-ion collisions:

(i) Initial condition.  HIJING model~\cite{Wang:1991hta,Gyulassy:1994ew} serves as the initial condition for the AMPT model and provides the spatial and momentum information of minijet partons and soft string excitations.
In order to increase the simulation efficiency for jet quenching study, the dijet production is triggered with the help of the jet triggering technique in HIJING model~\cite{Wang:1991hta,Gyulassy:1994ew}, which produces a triggered dijet with a specified $p_{T}$ in each event.
Several hard QCD processes are taken into account in the triggered dijet production: $qq \rightarrow qq$, $q\bar{q}\rightarrow q\bar{q}$, $q\bar{q}\rightarrow gg$, $qg\rightarrow qg$, $gg\rightarrow q\bar{q}$, and $gg\rightarrow gg$.
All initial-state and final-state radiation corrections are included in the AMPT model, therefore, a high-$p_{T}$ primary parton evolves into a full jet shower parton with lower virtualities.
In the string melting version of the AMPT model~\cite{Lin:2004en}, the triggered jets and minijets are first combined with their parent strings to form excited strings which fragment into hadrons according to the Lund string fragmentation~\cite{Sjostrand:2000wi}.
Then all hadrons are converted back into quarks and anti-quarks according to the flavor and spin structures of their valence quarks, forming the parton plasma.

(ii) Parton cascade. The dynamical evolution of the parton plasma is simulated by Zhang's parton cascade (ZPC) model~\cite{Zhang:1997ej}, which describes elastic partonic collisions among the medium partons and jet partons. The interaction strength of the elastic collisions is controlled by the partonic cross section $\sigma$, which is further determined by the value of the strong coupling constant and the Debye screening mass.

(iii) Hadronization. When the collisions of all partons stop, the AMPT model hadronizes all partons via a simple coalescence model which combines two nearest quarks into a meson and three nearest quarks into a baryon.

(iv) Hadronic rescattering. The dynamics of the subsequent hadronic interactions is then simulated via a relativistic transport (ART) model~\cite{Li:1995pra} which includes baryon-baryon, baryon-meson, meson-meson elastic and inelastic scatterings.

In this work, we simulate Pb$+$Pb collisions at 2.76~ATeV using the parameters that have been fitted to describe the soft bulk observables at the LHC energies~\cite{Xu:2011fi}.
Three sets of parton interaction cross section (0, 1.5~mb, 3.0~mb) are used to simulate the jet evolution in the dense partonic matter created in Pb$+$Pb collisions: $\sigma=0$ is to mimic the scenarios with only hadronic interactions (pp collisions), while $\sigma=1.5$~mb and $3.0$~mb is to investigate the effect from different partonic interaction strengths.

\begin{widetext}

\begin{figure}[tbh]
	\includegraphics[scale=0.67]{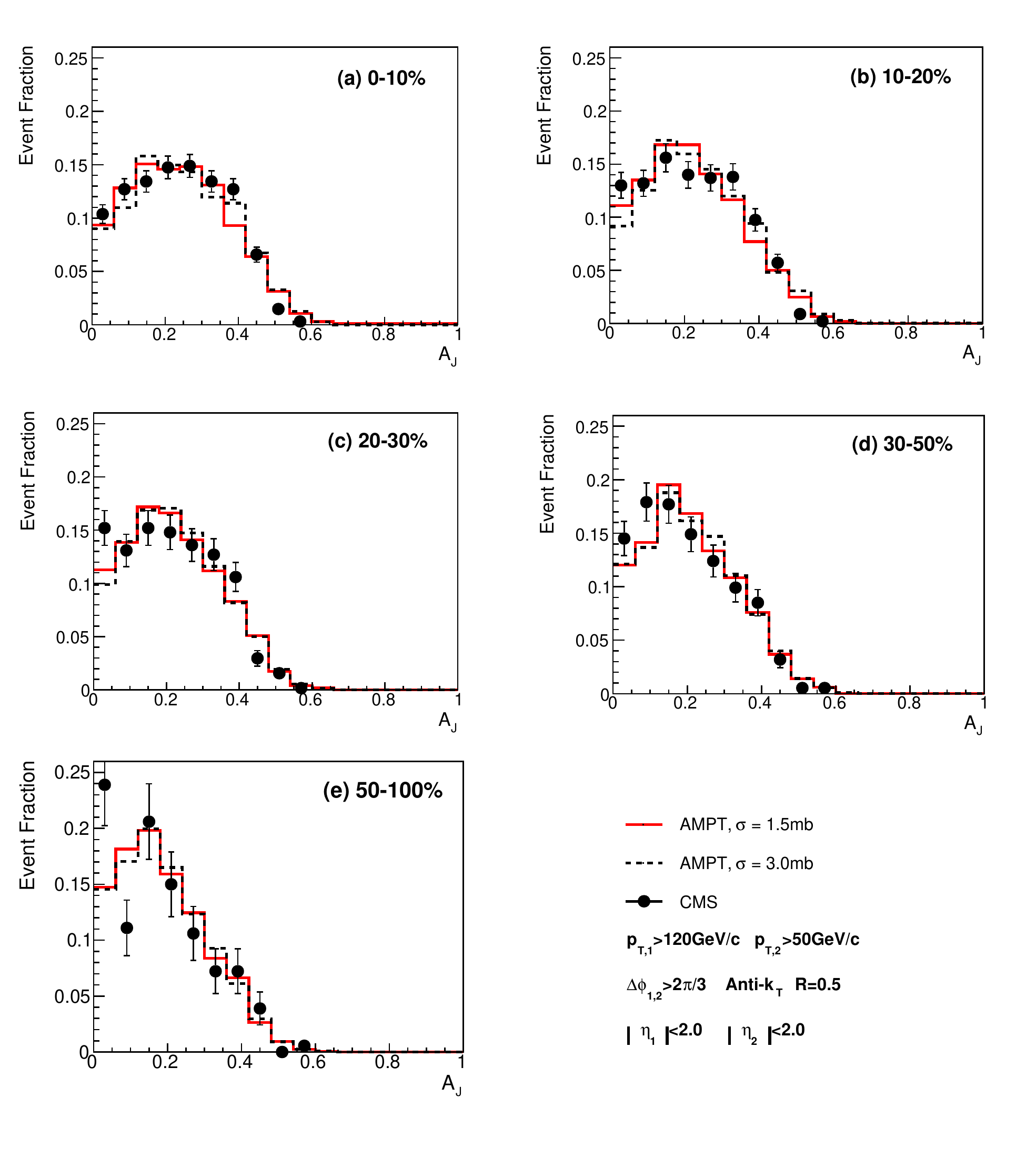}
	\caption{Dijet $A_{J}$ event distributions in 0-10\%, 10-20\%, 20-30\%, 30-50\% and 50-100\% PbPb collisions at 2.76~ATeV from the AMPT model ($\sigma$ = 1.5 and 3.0 mb), compared to CMS data~\cite{Chatrchyan:2011sx}. The cone size for jet reconstruction is $R=0.5$.}
	\label{fig:Aj}
\end{figure}

\begin{figure}[tbh]
	\includegraphics[scale=0.67]{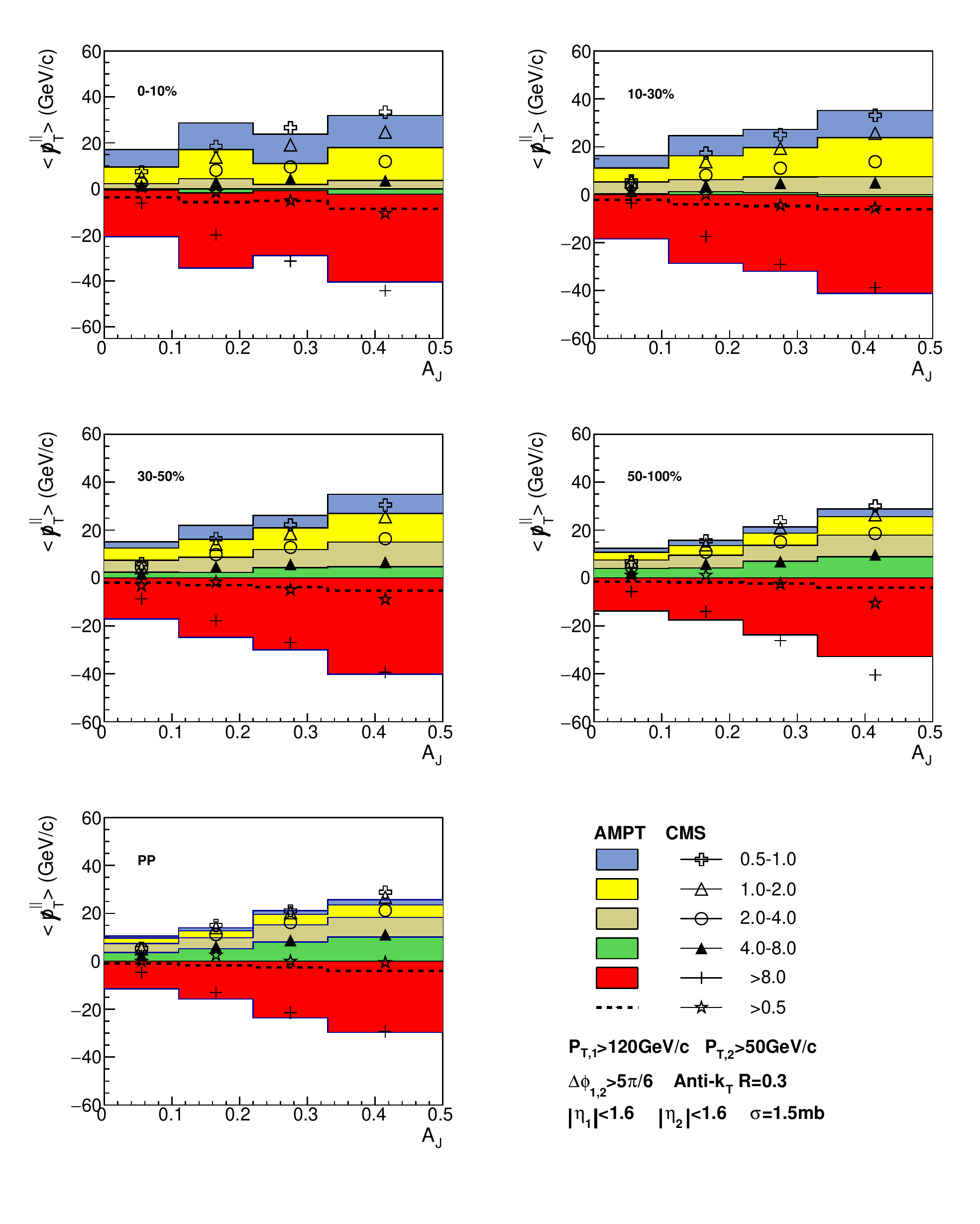}
	\caption{The event-averaged transverse momentum projection $\langle \slashed{p}_{T}^{||} \rangle$ calculated from the AMPT model as a function of $A_J$ for pp collisions, and 50-100\%, 30-50\%, 10-30\%, 0-10\% PbPb collisions at 2.76~ATeV.
Each band shows the contribution from charged hadrons with $p_T=$0.5-1~GeV/$c$, 1-2~GeV/$c$, 2-4~GeV/$c$, 4-8~GeV/$c$, $p_{T}>8.0$~GeV/$c$.
The solid lines at the edges of the bands denote the cumulative contributions from different $p_T$ ranges, and the thick dashed line denotes the cumulative contribution from charged hadrons with $p_{T}>0.5$~GeV/$c$.
The symbols represent the CMS data ~\cite{CMS:2014uca} for cumulative contributions from different $p_T$ ranges (to be compared to the corresponding lines calculated from the AMPT model).
The cone size for jet reconstruction is $R=0.3$. The partonic cross section $\sigma=1.5$~mb.
}
	\label{fig:pslash}
\end{figure}

\begin{figure}[tbh]
	\includegraphics[scale=0.67]{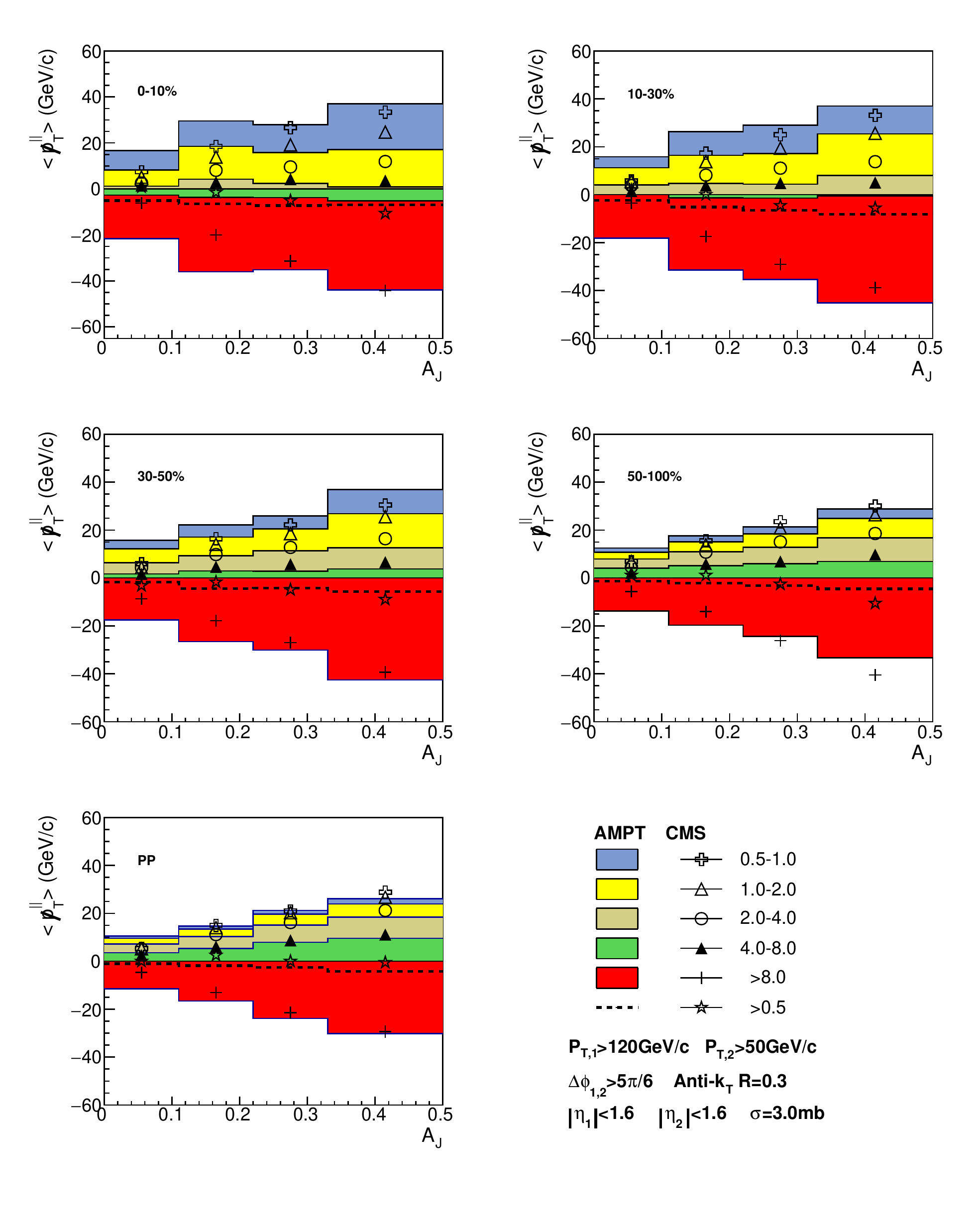}
	\caption{Same as Fig.~\ref{fig:pslash} but for $\sigma = 3.0$~mb.
	}
\label{fig:pslash3.0}
\end{figure}

\begin{figure}[tbh]
	\includegraphics[scale=0.67]{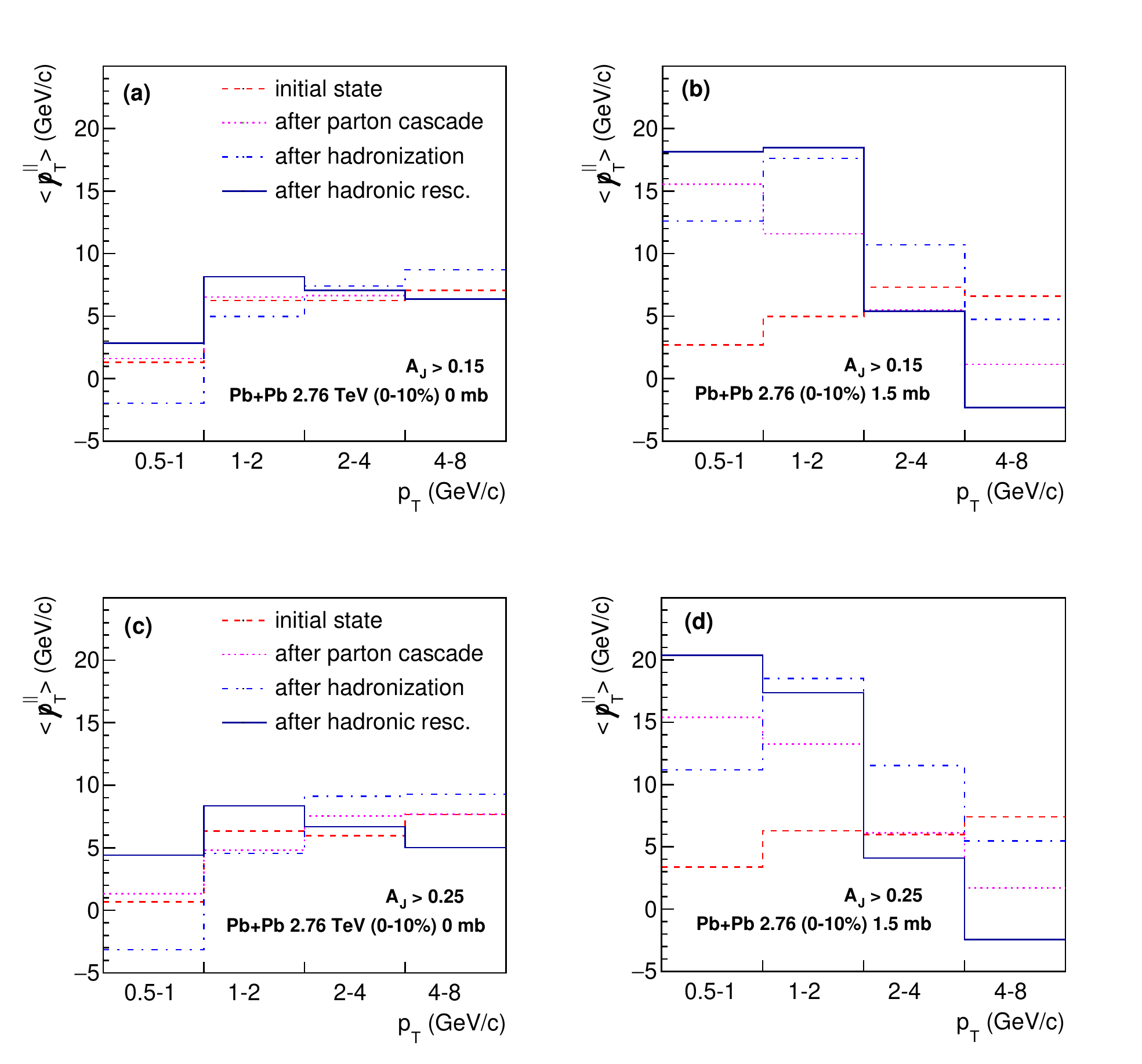}
	\caption{$\langle \slashed{p}_{T}^{||} \rangle$ as a function of $p_T$ bin in different evolution stages for 0-10\% PbPb collisions at 2.76~ATeV calculated from the AMPT model, where the dijet asymemetry is taken as $A_J > 0.15$ (upper) and $A_J > 0.25$ (lower), and the partonic cross section is taken to be 0 mb (left) and 1.5 mb (right).}
		\label{fig:pts_df}
\end{figure}

\begin{figure}[tbh]
	\includegraphics[scale=0.67]{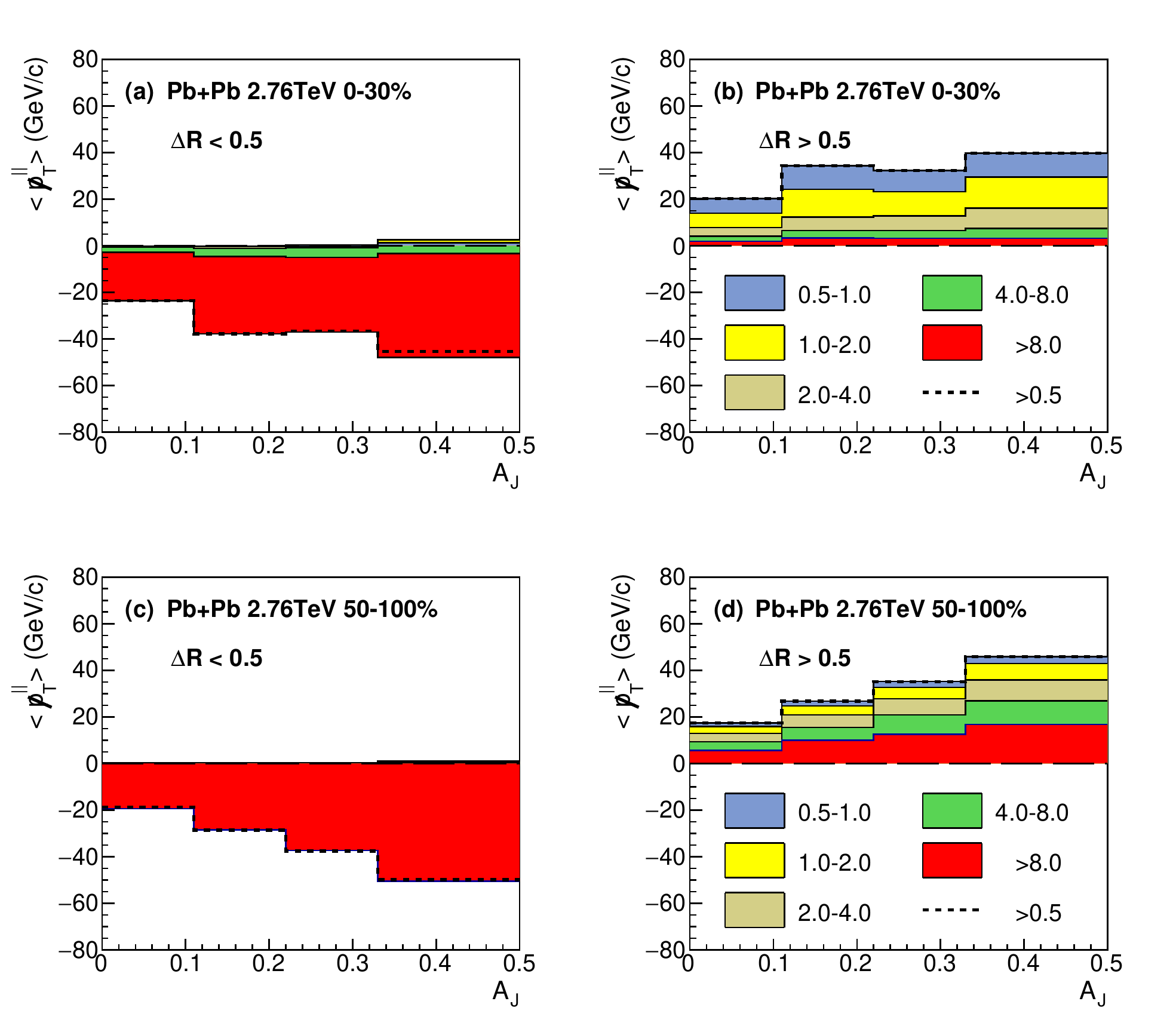}
	\caption{The in-cone (left) and out-of-cone (right) contributions to the event-averaged transverse momentum projection $\langle \slashed{p}_{T}^{||} \rangle$ as a function of $A_{J}$ in 0-30\% (upper) and 50-100\% (lower) PbPb collisions at 2.76~ATeV calculated from the AMPT model ($\sigma=1.5$~mb). The cone size is $0.5$.}
	\label{fig:inout0p5}
\end{figure}

\begin{figure}[tbh]
	\includegraphics[scale=0.67]{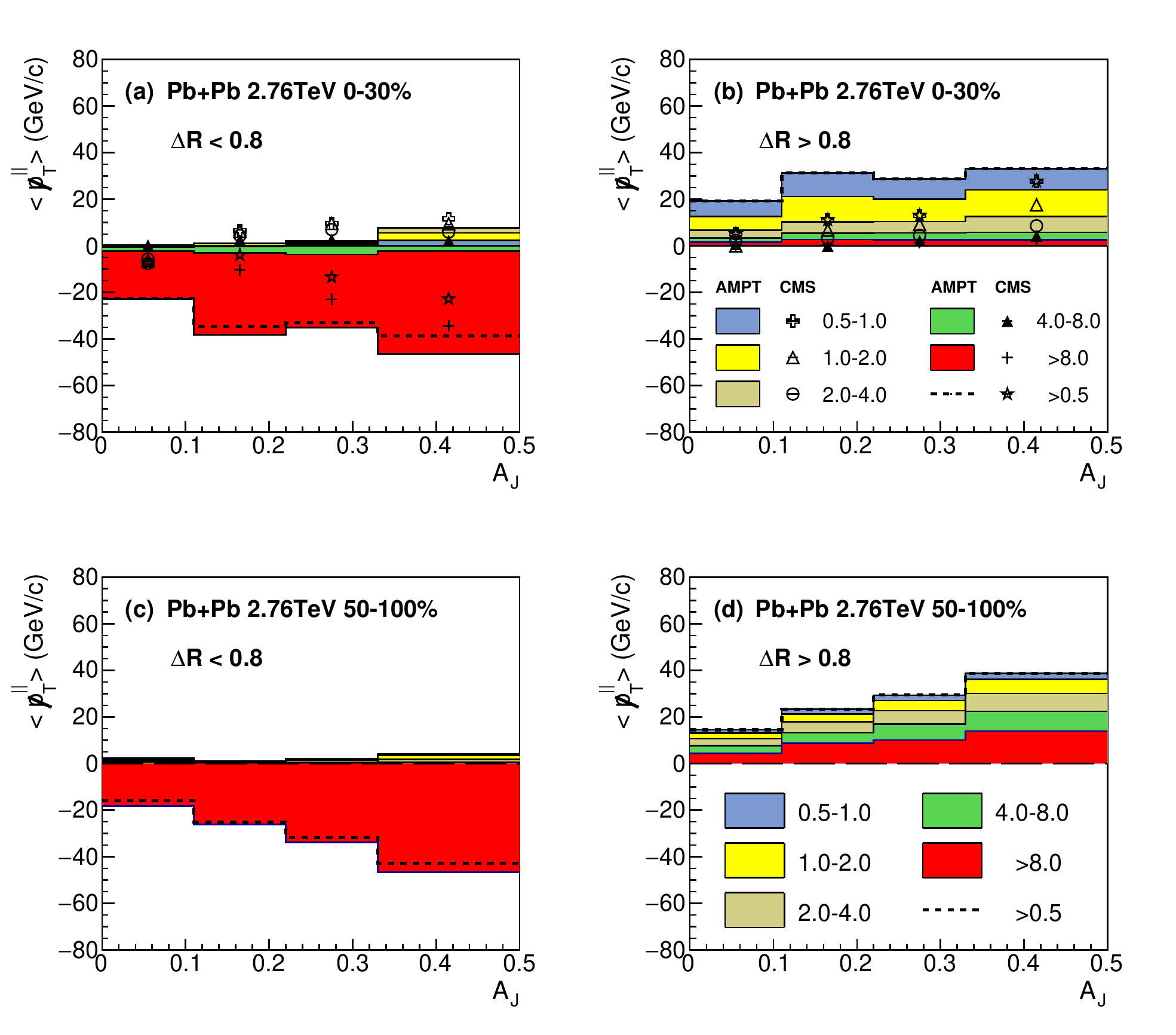}
	\caption{Same as Fig.~\ref{fig:inout0p5} but for cone size $0.8$. The CMS data is from the reference ~\cite{Chatrchyan:2011sx}.  }
	\label{fig:inout0p8}
\end{figure}

\begin{figure}[tbh]
	\includegraphics[scale=0.67]{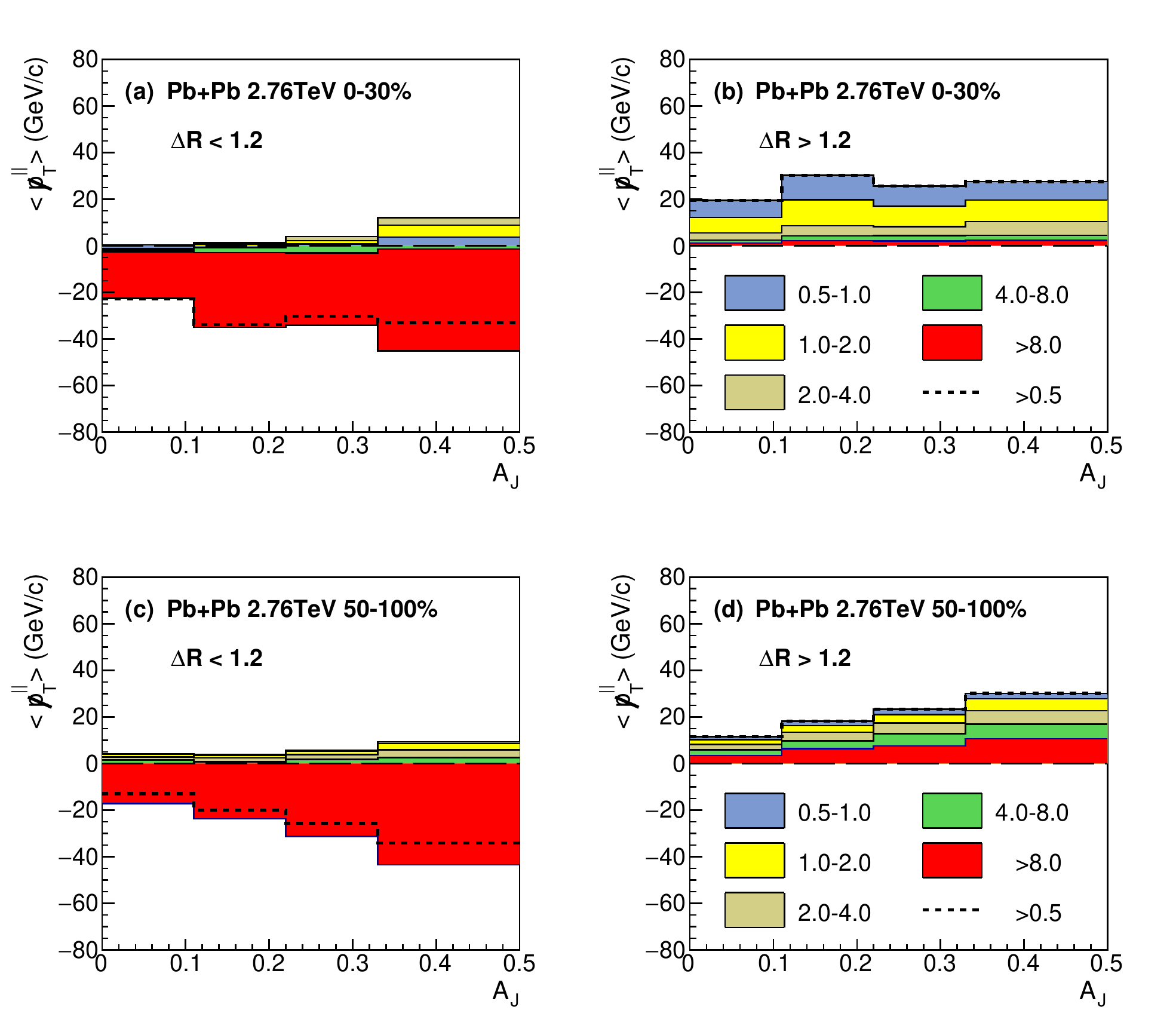}
	\caption{Same as Fig.~\ref{fig:inout0p5} but for cone size $1.2$.}
	\label{fig:inout1p2}
\end{figure}

\end{widetext}

\section{Results}
\label{sec:result}

To study full jets with the AMPT model, we utilize the standard Fastjet package~\cite{Cacciari:2011ma} with the anti-$k_{T}$ algorithm to reconstruct the full jets from the output of the AMPT simulation. To compare with the measurements by CMS Collaboration, we apply the same kinematics for jet cone sizes, transverse momenta, pseudorapidity cuts, and azimuthal angular cuts when reconstructing full jets and studying transverse momentum imbalance distribution and different contributions to the overall momentum balance of asymmetric dijet events.
We also simulate the effects of the background fluctuations and detector response via applying a Gaussian smearing to the jet $p_T$ obtained from AMPT; the centrality-dependent smearing widths $\sigma(p_T^{\rm obs}/p_T^{\rm AMPT})$ are taken from \cite{Chatrchyan:2011sx, CMS:2014uca}.

We first check the transverse momentum imbalance for asymmetric dijet events from the AMPT model by studying the asymmetry variable $A_J$ defined as follows:
\begin{equation}
A_{J} = \frac{p_{T,1} - p_{T,2}}{p_{T,1} + p_{T,2}},
\label{eqn:Aj}
\end{equation}
where the subscripts 1 and 2 denote the leading and the subleading jets, respectively.
The numerical results from the AMPT model simulation are shown in Fig.~\ref{fig:Aj}, compared with the CMS measurements~\cite{Chatrchyan:2011sx}.
Here the dijet momentum imbalance $A_J$ distributions (the event fractions) are plotted as a function of $A_{J}$ for PbPb collisions at 2.76~ATeV.
We apply the same kinematic cuts as the CMS measurements: the jet cone size $R=0.5$, the leading jet $p_{T,1}>$120 GeV/$c$, the subleading jet $p_{T,2}>$50 GeV/$c$, jet pseudorapidity cut $|\eta_{1,2}|<2$, and the relative azimuthal angle between the leading and subleading jets $\Delta \phi_{12} = |\phi_1 - \phi_2| > 2\pi/3$.
Five different centrality bins are shown: 0-10\%, 10-20\%, 20-30\%, 30-50\%, and 50-100\%.
Two different values for the partonic cross section are used: $\sigma=1.5$~mb and $3.0$~mb.

From Fig.~\ref{fig:Aj}, we can see that from the most peripheral to the most central PbPb collisions, the dijet momentum imbalance $A_J$ distribution shifts to the right (i.e., larger $A_J$ values).
This indicates the away-side subleading jets may experience a significant amount of energy loss due to the interactions with the bulk matter when they pass through the dense partonic medium produced in PbPb collisions.
In central collisions with denser and larger medium, jet-medium interaction is stronger, which leads to larger jet energy loss and thus larger asymmetry $A_J$.
The results from the AMPT model calculation agree well with the CMS dijet $A_J$ measurements, consistent with a previous study~\cite{Ma:2013pha}.

To further investigate where the lost energy from hard jets goes and how it is redistributed, we follow CMS collaboration~\cite{Chatrchyan:2011sx} and study the overall momentum balance for asymmetric dijet events.
One can project the transverse momenta $p_T$ of all the final charged hadrons onto the leading jet axis, i.e., for each event, the projected transverse momentum $\slashed{p}_{T}^{||}$ can be defined:
\begin{equation}
\slashed{p}_{T}^{||} = \sum_{i} - p_{T}^{i}\cos(\phi_{i} - \phi_{leading~jet}),
\label{eqn:pslash}
\end{equation}
where we take the sum over all final charged hadrons with transverse momenta $p_{T} > 0.5$ GeV/$c$ and pseudorapidity $\left| \eta \right| < 2.4$ in following calculations.
We then average over the transverse momentum projection $\slashed{p}_{T}^{||}$ over all simulated events for given $A_{J}$ bins.
According to the definition, the negative $\langle \slashed{p}_{T}^{||} \rangle $ denotes the projection of the transverse momentum in the direction of the leading jets, while the positive $\langle \slashed{p}_{T}^{||} \rangle$ represents the projection in the opposite direction of the leading jets.

The numerical results from the AMPT model calculation are shown in Fig.~\ref{fig:pslash} ($\sigma=1.5$~mb) and Fig.~\ref{fig:pslash3.0} ($\sigma=3.0$~mb) , where the event-averaged transverse momentum projection $\langle \slashed{p}_{T}^{||} \rangle $ is plotted as a function of the asymmetry variable $A_{J}$ for both pp and PbPb collisions at 2.76~TeV.
We use the same kinematic cut as the CMS analysis: the jet cone $R = 0.3$, the leading jet $p_{T,1}>$120 GeV/$c$, the subleading jet $p_{T,2}>$50 GeV/$c$, the pseudorapidity cut $|\eta_{1,2}|<1.6$, and the azimuthal angle cut $\Delta \phi_{12} > 5\pi/6$.
Four different centrality bins are shown: 0-10\%, 10-30\%, 30-50\% and 50-100\%.
For each centrality bin, we show the individual contributions to the projected transverse momentum $\langle \slashed{p}_{T}^{||} \rangle $ from five different $p_T$ regions: 0.5-1.0~GeV/$c$, 1.0-2.0~GeV/$c$, 2.0-4.0~GeV/$c$, 4.0-8.0~GeV/$c$ and $p_{T}>8.0$~GeV/$c$ (which are denoted by different bands, respectively, in each plot).
The solid curves at the edges of the bands show the cumulative contributions from the combinations of different $p_{T}$ bins, and the thick dashed curves denotes the total contribution from the sum of all charged hadrons with $p_{T} > 0.5$ GeV/$c$.
The CMS data are denoted by different symbols, which are then compared to the cumulative contributions from different $p_T$ bins from the AMPT model simulation (denoted by the edges of the bands).

From Fig.~\ref{fig:pslash}, we can see that in both pp and PbPb collisions, there is a large negative contribution (i.e., in the direction of the leading jets) to the transverse momentum projection $\langle \slashed{p}_{T}^{||} \rangle $, which is dominated by hard charged hadrons ($p_T>8.0$~GeV/$c$).
The imbalance contributed from hard hadrons increases from peripheral to central PbPb collisions, which might indicate that the subleading jets tend to hadronize into less hard fragments due to stronger jet-medium interaction and energy loss, as compared to the leading jets.
Such negative contribution from hard hadrons is mostly balanced by the combined positive contributions from the hadrons with $p_T=0.5$-8~GeV/$c$.
Therefore, the overall projected transverse momentum $\langle \slashed{p}_{T}^{||} \rangle_{\rm total}$ with all hadrons with $p_T > 0.5$~GeV/$c$ is roughly balanced in both pp and PbPb collisions, as required by the momentum conservation.
Due to the kinematic cuts applied on the transverse momentum ($p_T > 0.5$~GeV/$c$), the pseudorapidity ($|\eta| < 2.4$), etc., there is still some remaining transverse momentum imbalance after taking into account all the charged hadrons with $p_T > 0.5$~GeV/$c$ (shown by dashed curves).

Another important observation is that from pp (or peripheral PbPb) to central PbPb collisions, the positive contribution (in the opposite direction of the leading jets) to the projected transverse momentum $\langle \slashed{p}_{T}^{||} \rangle $ from the soft hadrons ($p_T = 0.5$-$2.0$~GeV/$c$) gradually increases and finally dominates in the most central PbPb collisions.
This shows that a large portion of the lost energy from the jets is carried by the final state soft hadrons.
We note that only elastic processes are included in the AMPT model, which might indicate that elastic scatterings could play significant roles in transporting the lost energy from hard jets into the soft hadrons.
The AMPT model describes quite well the CMS measurements of the overall momentum balance in asymmetric dijet events for both pp and peripheral PbPb collisions.
AMPT also gives a good description of the centrality dependence for the positive contribution from soft hadrons to the transverse momentum projection $\langle \slashed{p}_{T}^{||} \rangle$.

Comparing Fig.~\ref{fig:pslash3.0} with $\sigma=3.0$~mb to Fig.~\ref{fig:pslash} with $\sigma=1.5$~mb, we can see that when the partonic cross section increases, the positive contribution from the soft hadrons ($p_T = 0.5$-$2.0$~GeV/$c$) also increases, which is more visible for more central PbPb collisions. The reason is that with larger parton cross section, there is more jet energy loss due to stronger jet-medium interaction, thus more energies is transported outside jet cone and carried by the final state soft hadrons.

To trace back how the difference on the overall momentum balance for asymmetric dijet events in pp collisions and central PbPb collisions is developed, we calculate the positive contributions to the projected transverse momentum $\langle \slashed{p}_{T}^{||} \rangle $ for four different evolution stages in the AMPT model: (i) initial state jet production from HIJING, (ii) after parton cascade, (iii) after hadronization, (iv) after hadron rescattering.
The numerical result is shown in Fig.~\ref{fig:pts_df}, where $\langle \slashed{p}_{T}^{||} \rangle$ for partons (in the 1st and 2nd stages) and hadrons (in the 3rd and 4th stages) with $p_T = 0.5$-$8.0$~GeV/$c$ is plotted as a function of $p_T$ bin for both pp and central PbPb collisions.
Here we show the results for two different dijet asymmetry cuts $A_{J}>0.15$ (upper) and $A_{J}>0.25$ (lower) with two different values of cross section $\sigma=0$ (to mimic pp collisions) and $1.5$~mb.

One can see that in central PbPb collisions~($\sigma = 1.5$~mb), the relative contributions from different $p_T$ particles to the transverse momentum projection $\langle \slashed{p}_{T}^{||} \rangle$ changes dramatically after parton cascade, in contrast to the result in PbPb collisions with $\sigma = 0$~mb (or pp collisions, not shown).
More specifically, the contribution from soft particles to the transverse momentum projection $\langle \slashed{p}_{T}^{||} \rangle$ increases while the contribution from high $p_T$ particles decreases.
This indicates that jet-medium interaction in the partonic stage (elastic scatterings in the AMPT model) may give important contribution to the transport (redistribution) of the lost energy from the jets.
We note that the hadronization process and hadronic interaction in the AMPT model also give visible contributions.
Compared to the partonic interaction, the hadronic rescattering generates similar though a little smaller effect on $\langle \slashed{p}_{T}^{||} \rangle$.
Notably, the recombination mechanism that converts partons into hadrons produces the opposite effect: the contribution from lower $p_T$ particles decreases while the contribution from higher $p_T$ particles increases.
Also by comparing the results with two different $A_J$ cuts, we find that the above modifications are stronger for $A_J=0.25$ than $A_J=0.15$ (as a consequence of larger jet energy loss).

Following CMS Collaboration, we further study the angular distribution of the lost energy carried by the final state soft hadrons by dividing the total contributions to the projected transverse momentum  $\langle \slashed{p}_{T}^{||} \rangle $ into two angular regions: one from the charged hadrons inside the cone $\Delta R = \sqrt{(\phi - \phi_{J})^2 + (\eta - \eta_{J})^2}$ around the leading jet axis or the opposite direction, the other is outside the cone $\Delta R$, where $\phi$ and $\eta$ are the azimuthal angle and pseudorapidity of the charged hadrons, and $\phi_J$ and $\eta_J$ are the azimuthal angle and pseudorapidity of the leading jet or sub-leading jet, respectively.
With increasing cone size $\Delta R$, more hadrons are included in the in-cone contribution and excluded from the out-of-cone contribution.
For very large $\Delta R$, one includes all charged hadrons in the cones, then the in-cone contribution reduces to the result in Fig.~\ref{fig:pslash}.

In Fig.~\ref{fig:inout0p5}, \ref{fig:inout0p8} and \ref{fig:inout1p2}, we show the numerical results the in-cone (left) and out-of-cone (right) contributions to the event-averaged transverse momentum projections  $\langle \slashed{p}_{T}^{||} \rangle $ as a function of the asymmetry variable $A_{J}$, for central (upper) and peripheral (lower) PbPb collisions at 2.76~ATeV, and for three different cone sizes $\Delta R=$ $0.5$, $0.8$ and $1.2$, respectively.
One can clearly see that in both central and peripheral PbPb collisions, and also for three different cone size $\Delta R$ values, the in-cone contribution to the projected transverse momentum  $\langle \slashed{p}_{T}^{||} \rangle $ is dominated by large $p_T$ hadrons ($p_T>8.0$~GeV/$c$) which come from the hard fragments of the reconstructed leading and subleading jets.
The soft hadrons give quite small in-cone contribution, but the contribution increases when moving from peripheral to central PbPb collisions or increasing the cone size $\Delta R$.
For the out-of-cone contribution to $\langle \slashed{p}_{T}^{||} \rangle $,
we can see that for peripheral PbPb collisions, there is still a sizable contribution from large $p_T$ hadrons to $\langle \slashed{p}_{T}^{||} \rangle $.
However, for central PbPb collisions, the out-of-cone contribution becomes more dominated by the soft hadrons ($p_T = 0.5-2.0$~GeV/$c$).
This indicates that a large fraction of the momentum imbalance in asymmetric dijet events, originating primarily from jet-medium interactions and jet energy loss in the partonic stage, is balanced by the soft hadrons at large angles away from the dijet axis.
Note that the AMPT model only includes elastic processes in the parton cascade, which might indicate that elastic collisions could contribute significantly to the redistribution of the lost energy from hard jets into soft hadrons emitted at large angles away from the dijet axis.

In addition, it is interesting to observe that compared to CMS data in Fig. \ref{fig:inout0p8}, the AMPT simulation overestimates the individual positive and negative contributions to the overall transverse momentum balance.
A possible reason for this overestimation is the neglect of the radiative processes in the AMPT model: to obtain similar amount of jet energy loss and dijet asymmetry, we have included more contributions from elastic collisions than would be required when radiative processes are included.
Given that the medium-induced radiation is more dominated by the collinear phase space, elastic scatterings should be more effective in transporting the momentum to the directions transverse to the jet axis.
Therefore, without the radiative processes in the AMPT model, relatively more of the lost energy is transported to the final state soft hadrons emitted at large angles from the jet axis (via elastic scatterings).
Further study utilizing both radiative and collisional processes should be helpful to clarify this issue, which we leave as a future effort.

\section{SUMMARY}
\label{sec:sum}

Within the framework of the AMPT model, we have studied the overall transverse momentum balance and the redistribution of the lost energy from hard jets for asymmetric dijet events in PbPb (and pp) collisions at 2.76~ATeV at the LHC.
In particular, we have performed a detailed analysis of the projected transverse momentum $\langle \slashed{p}_{T}^{||} \rangle$ contributed from the final-state charged hadrons which carry different transverse momenta and are emitted at different angular directions with respect to the dijet axis.

For the overall transverse momentum balance, we found that the large negative contribution in the direction of the leading jets to the projected transverse momentum $\langle \slashed{p}_{T}^{||} \rangle $ is dominated by hard hadrons ($p_T>8.0$~GeV/$c$) for both peripheral and central PbPb collisions.
In contrast, soft hadrons ($p_T = 0.5$-$2.0$~GeV/$c$) contribute to positive $\langle \slashed{p}_{T}^{||} \rangle $.
The positive contribution from soft hadrons increases with increasing collision centrality, and dominates in the most central PbPb collisions.
This suggests that a large fraction of the lost energy from hard jets is carried by the final state soft hadrons.
We have further calculated the positive contributions to  $\langle \slashed{p}_{T}^{||} \rangle $ in each evolution stage in the AMPT model which showed that the increasing soft-hadron contribution is mainly developed in the parton cascade (though hadronization and hadronic interaction also give sizable contribution), and elastic collisions can effectively transport the lost energy from jets to partons which are fragmented into soft hadrons.

We have also investigated the redistribution of the lost energy in the angular direction by dividing the overall momentum balance into in-cone and out-of-cone contributions relative to the dijet axis.
It was found that the in-cone contribution to the projected transverse momentum  $\langle \slashed{p}_{T}^{||} \rangle $ is dominated by large $p_T$ hadrons ($p_T>8.0$~GeV/$c$) in both central and peripheral PbPb collisions, also for three different cone sizes ($\Delta R = 0.5, 0.8, 1.2$).
For the out-of-cone contribution to $\langle \slashed{p}_{T}^{||} \rangle $, while there is still a sizable contribution from large $p_T$ hadrons in peripheral PbPb collisions, soft hadrons dominate in the most central PbPb collisions.
Since the AMPT model only includes elastic processes, the qualitative agreement of our result with the CMS data might indicate that the elastic collisions could play important roles in the transportation of the lost energy from the hard jets to very large angles.
Our present study constitutes an important contribution to our understanding of the interaction between hard jets and medium.
Future studies with radiative processes will provide additional insight into this issue.

\section{ACKNOWLEDGMENTS}
\label{sec:ack}

This work was
supported by the Natural Science Foundation of China (NSFC) under Grant No.~11435004 and No.~11375072, and by the Major State Basic Research Development Program in China under Grant No.~2014CB845400.
G.-L. Ma is supported by NSFC under Grants No.~11522547, No.~11375251, and No.~11421505, and by the Youth Innovation Promotion Association of CAS under Grant No.~2013175.


\bibliographystyle{h-physrev5}
\bibliography{refs_GYQ}

\end{document}